\documentclass[a4paper,11pt]{article}
\pdfoutput=1 
\usepackage[utf8]{inputenc} 
\usepackage{jheppub} 

\usepackage{graphicx}  
\usepackage{dcolumn}   
\usepackage{bm,relsize}        
\usepackage{amssymb,amsmath,mathrsfs}
\usepackage{textcomp}
\usepackage{wasysym}
\usepackage{slashed}
\usepackage{multirow}
\usepackage{lipsum, color}
\usepackage[usenames,dvipsnames,svgnames]{xcolor}
\usepackage{booktabs}
\usepackage{xfrac}
\usepackage[capitalize]{cleveref}

\def\beq{\begin{equation}}
\def\eeq{\end{equation}}
 \def\be{\begin{equation}} \def\ee{\end{equation}}
\def\bea{\begin{eqnarray}} \def\eea{\end{eqnarray}}

\newcommand{\hdark}{{h_d}}
\newcommand{\Adark}{{\gamma_d}}

\newcommand{\radiuslayerone}{23.3~mm}
\newcommand{\radiuslayertwo}{29.8~mm}
\newcommand{\radiuslayerthree}{73.9~mm}
\newcommand{\radiuslayerfour}{86.3~mm}

\newcommand{\lengthlayerone}{124.7~mm}
\newcommand{\lengthlayertwo}{124.7~mm}
\newcommand{\lengthlayerthree}{351.9~mm}
\newcommand{\lengthlayerfour}{372.6~mm}

\newcommand{\mufiveeSMfull}{\mu^+ \to e^+e^+e^-e^+e^-\overline{\nu}_\mu \nu_e}
\newcommand{\mufiveeSM}{\mu^+ \to 3e^+\,2e^-\,2\nu}
\newcommand{\mufiveefull}{\mu^+ \to e^+e^+e^-e^+e^-}
\newcommand{\mufivee}{\mu^+ \to 3e^+\,2e^-}
\newcommand{\muthreeeSMfull}{\mu^+ \to e^+ e^+ e^-\overline{\nu}_\mu \nu_e}
\newcommand{\muthreeefull}{\mu^+ \to e^+ e^+ e^-}
\newcommand{\muthreee}{\mu^+ \to 2e^+\,1e^-}

\usepackage{fontawesome5} 
\definecolor{blue-violet}{rgb}{0.33, 0.17, 0.89}

\begin{document}

\title{New physics in multi-electron muon decays}

\author[a,b,c]{Matheus Hostert,}
\author[d]{Tony Menzo,}
\author[b,c]{Maxim Pospelov,}
\author[d]{Jure Zupan,}
\affiliation[a]{Perimeter Institute for Theoretical Physics, Waterloo, ON N2J 2W9, Canada}
\affiliation[b]{School of Physics and Astronomy, University of Minnesota, Minneapolis, MN 55455, USA}
\affiliation[c]{William I. Fine Theoretical Physics Institute, School of Physics and Astronomy, University of Minnesota, Minneapolis, MN 55455, USA}
\affiliation[d]{Department of Physics, University of Cincinnati, Cincinnati, Ohio 45221, USA}

\emailAdd{mhostert@pitp.ca}
\emailAdd{menzoad@mail.uc.edu}
\emailAdd{pospelov@umn.edu}
\emailAdd{zupanje@ucmail.uc.edu}

\date{\today}

\abstract{
We study the exotic muon decays with five charged tracks in the final state.
First, we investigate the Standard Model rate for $\mufiveeSM$ ($\mathcal{B}=4.0\times 10^{-10})$ and find that the Mu3e experiment should have tens to hundreds of signal events per $10^{15}$ $\mu^+$ decays, depending on the signal selection strategy.
We then turn to a neutrinoless $\mufivee$ decay that may arise in new-physics models with lepton-flavor-violating effective operators involving a dark Higgs $h_d$.
Following its production in $\mu^+ \to e^+ h_d$ decays, the dark Higgs can undergo a decay cascade to two $e^+e^-$ pairs through two dark photons, $h_d \to \gamma_d \gamma_d \to 2(e^+e^-)$.
We show that a $\mufivee$ search at the Mu3e experiment, with potential sensitivity to the branching ratio at the ${\mathcal O}(10^{-12})$ level or below, can explore new regions of parameter space and new physics scales as high as $\Lambda \sim 10^{15}$~GeV.
}

\maketitle

\section{Introduction}

The search for charged lepton flavor violation (cLFV) in ultra-rare decays is achieving experimental sensitivity to new physics scales far above the TeV scale.
Among the different flavor-violating observables, muon decays stand out as a particularly clean and sensitive probe --- muons can be copiously produced and easily manipulated in the laboratory.
The Mu3e experiment, under construction at the Paul Scherrer Institut (PSI), will be sensitive to $\mathcal{B}(\muthreeefull)$ as small as $10^{-16}$~\cite{Mu3e:2020gyw}, significantly improving on existing limits from SINDRUM, at the level of $\mathcal{B}(\muthreeefull) < 10^{-12}$~\cite{SINDRUM:1987nra}.
This indirectly probes cLFV at new-physics scales as large as $\Lambda \sim 10^{4}$~TeV.
Mu3e is also complementary to many other muon facilities, including searches for $\mu^+ \to e^+ \gamma$ at MEG and MEG-II~\cite{MEG:2016leq,MEGII:2018kmf}, as well as $\mu \to e$ conversion on nuclei at COMET~\cite{COMET:2018auw}, DeeMe~\cite{Teshima:2019orf}, and Mu2e~\cite{Mu2e:2014fns}.

The Mu3e detector is a magnetized thin spectrometer designed to contain and track as many electrons and positrons from stopped muons as possible.
The low density of each active layer reduces backgrounds from particle interactions in the material and minimizes the energy loss of charged tracks, thereby improving momentum resolution.
The latter is an important capability to identify missing energy in muon decays and reduce the Standard Model (SM) backgrounds from, e.g., $\muthreeeSMfull$.
Due to the sheer number of muon decays and the precision required for a low-background environment, Mu3e can also be useful for applications beyond cLFV.
For instance, Mu3e can be used to perform precision measurements of the SM decays $\muthreeeSMfull$ that require improved precision of theoretical predictions~\cite{Fael:2016yle,Pruna:2016spf,Banerjee:2020rww} or to perform searches for exotic new particles~\cite{Echenard:2014lma,Flores-Tlalpa:2015vga,Heeck:2017xmg,Perrevoort:2018okj,Banerjee:2022nbr}.
This work proposes a new auxiliary goal for the experiment: searches for the more exotic but clean decay channels $\mufiveeSMfull$ and $\mufiveefull$.

We assess the Mu3e sensitivity to the SM decay $\mufiveeSMfull$, which we find to have the branching ratio of $4.0\times 10^{-10}$. 
Such a process can be of interest in its own right as one of the rarest radiative processes in the decay of the 2$^{\rm nd}$ and 3$^{\rm rd}$ generation particles that any experiment can directly observe.
It is also a background to the new physics searches in neutrinoless decays, $\mufivee$, but we show that it can be eliminated with an appropriate cut on missing momentum.

Rare decays have been a successful avenue to constrain light dark sectors; see, e.g., the reviews in Refs.~\cite{Goudzovski:2022vbt,Antel:2023hkf}.
Most of the experimental activity is focused on signatures where the new particle is reconstructed as a single visible or invisible resonance.
However, if the dark sector particle decays through a fast-developing cascade, multiple visible resonances may be present at no additional penalty to the original production rate.
While such scenarios come at the cost of non-minimality, they can be constrained more effectively due to their exotic nature.
Examples of such dark sectors have been studied in various contexts before, see for instance Refs.~\cite{Batell:2009yf,Batell:2009jf,Weihs:2011wp,Curtin:2014cca,Blinov:2017dtk,Smolkovic:2019jow,Hostert:2020xku,Foguel:2022unm,Ferber:2023iso} for phenomenological studies, and Refs.~\cite{BaBar:2012bkw,KLOE-2:2015nli,Belle:2020the,CMS:2021sch,ATLAS:2021ldb,CMS:2021pcy} for experimental searches involving multi-leptons.
In a similar model to the one discussed in this paper, Ref.~\cite{Hostert:2020xku} proposed searches for kaon decay modes with five charged tracks, $K^+ \to \pi^+ e^+ e^- e^+ e^-$, recently evaluated in the SM in Ref.~\cite{Husek:2022vul} and searched for at the NA62 experiment, reaching branching ratios as small as $\mathcal{O}(10^{-9})$~\cite{na62search}.

In this article, we turn to the $5e$ lepton-flavor-violating muon decay, {\em i.e.}, the $\mufiveefull$ process. 
To that end, we formulate the simplest dark sector model where the production of a light scalar takes place due to its cLFV interactions with the muon and electron.
As the dark Higgs of a $U(1)_d$ gauge group, this scalar decays to pairs of dark photons $\gamma_d$. 
Due to kinetic mixing with the SM, the dark photon subsequently decays into $e^+e^-$ pairs.
We analyze the experimental sensitivity to this mode with a simple Monte-Carlo (MC) and estimate the signal efficiency attainable at the Mu3e experiment. 
We conclude that a sensitivity to branching ratios as small as ${\mathcal O}(10^{-12})$ can be reached, constituting the most precise test of this class of dark sector models. 

The article is structured as follows. 
In \cref{sec:SMdecays} we present our calculations of the SM muon decay to five charged tracks.
We then discuss the dark sector models and associated five-track signatures in \cref{sec:exoticdecays}.
The experimental simulation and resulting sensitivity are presented in \cref{sec:experimentalreach}, and the conclusions in \cref{sec:concl}.

\section{Standard Model decays: $\mufiveeSMfull$}
\label{sec:SMdecays}

The enormous statistics expected to be collected by the upcoming muon experiments will enable the study of tiny radiative muon decay modes. 
These decay channels have important implications in their own right. 
Apart from testing QED at high orders in $\alpha$, they can also be important search channels for certain dark sector models. 
In what follows, we discuss the muon decay to five tracks with missing energy, i.e., muon decays with double internal photon conversions.

We estimate the branching ratios for $\mu \rightarrow 3e \bar{\nu}\nu$ and $\mu \rightarrow 5e \bar{\nu}\nu$ decays using \textsc{MadGraph5\_aMC@NLO}~\cite{Alwall:2014hca} to leading order in $G_F$ and $\alpha$.
We find
\begin{align}
\label{eq:SM3e}
\mathcal{B}(\muthreeeSMfull) &= (3.601 \pm 0.005) \times 10^{-5},
\\
\mathcal{B}(\mufiveeSMfull) &= (3.929 \pm 0.001) \times 10^{-10},
\label{eq:SM5e}
\end{align}
where the quoted uncertainty is from the MC only.
The electron mass is included to avoid infrared divergences, but the uncertainties in the masses and couplings are not taken into account.\footnote{The SM input parameters are $m_\mu = 105.658$~MeV, $m_e = 510.999$~keV, $\alpha^{-1}(\mu = 0) = 137.036$, $G_F =  1.16638 \times 10^{-5}$~GeV$^{-2}$, and $ \Gamma_\mu = 2.99598 \times 10^{-16}$~MeV~\cite{ParticleDataGroup:2022pth}.}
The $\muthreeeSMfull$ process has already been calculated previously~\cite{Fael:2016yle,Pruna:2016spf,Banerjee:2020rww} since it is an important background to the main physics goal of the Mu3e experiment, the search for neutrinoless $\mu^+\to e^+e^+e^-$ decays. 
It is also of interest in connection with the $\mu^+\to \overline\nu_\mu \nu_e e^+\gamma_d \to \overline\nu_\mu \nu_e e^+e^+e^-$ dark sector mode, where an additional vector particle ({\em e.g.} the dark photon) is radiated in the decay process \cite{Echenard:2014lma}. 
The analysis of sensitivity to $\gamma_d$ has been performed by the Mu3e collaboration~\cite{Perrevoort:2018okj}.
In addition, it is expected that the SM rate, \cref{eq:SM3e}, previously studied in \cite{SINDRUM:1985vbg}, will be observed with enormous statistics. 

The SM 5$e$ mode, \cref{eq:SM5e}, has not been calculated/discussed before, to the best of our knowledge. The rate is very small, but perhaps not hopeless with the statistics planned to be collected by Mu3e. 
The $\mu \to 5e2\nu$ decay occurs at order $\Gamma \propto {\mathcal O}(G_F^2 \alpha^4)$, where $G_F$ is the Fermi constant. 
The corresponding partial decay width, $\Gamma\sim {\mathcal O}(10^{-19}\,{\rm eV})$, would be the smallest measured decay rate involving the second and third-generation particles. 

To be observable, the five charged tracks should have sufficient energy to escape the Mu3e target and fall within the detector acceptance.
A simple estimate of the observable branching ratio can be found by requiring the transverse momentum of each electron and positron to be larger than the experimental threshold, $p_{\rm T, th} \sim 10$~MeV.
We find
\begin{align}\label{eq:SMBR_five_tracks}
    \mathcal{B}\left(\mufiveeSMfull\, |\,\text{all } p_{\rm e^\pm}^{\rm T, true} > 10\text{ MeV}\right) &= (1.4 \pm 0.1) \times 10^{-14},
\end{align}
where $p_{\rm e^\pm}^{\rm T, true}$ is the true transverse momentum of individual electrons and positrons in the muon rest frame.
Here ``true" refers to the value of the quantity before any experimental smearing is applied, and should be contrasted with the ``reconstructed" quantities on which we apply cuts in \cref{sec:experimentalreach}.
For completeness, we also quote the branching ratio for a threshold of $10$~MeV on the true \emph{total} momentum,
\begin{align}\label{eq:SMBR_ptot}
    \mathcal{B}\left(\mufiveeSMfull\, |\,\text{all } p_{\rm e^\pm}^{\rm true} > 10\text{ MeV}\right) &= (9.5 \pm 0.2) \times 10^{-13},
\end{align}
where $p_{e^\pm}^{\rm true}$  is the true momentum of individual electrons and positrons in the muon rest frame.
With the transverse-momentum cut, the above indicates that Mu3e should observe about $35$ events in 300 days of continuous running at a rate of $10^{8}$ muon stops per second, {\em i.e.}, about $2.5 \times 10^{15}$ muon decays.
We revisit these estimates in \cref{par:SMsensitivity} with our simulation, showing that requiring five observable tracks leads to an even smaller branching ratio.

Another relevant regime of the SM rate is when the neutrinos carry a small amount of energy.
Such phase-space configurations are highly disfavored in this decay since internal photon conversion tends to produce soft $e^+e^-$ pairs.
Nevertheless, it can represent a background to neutrinoless decay modes, as discussed in more detail in \cref{sec:backgrounds}.
The rate is a steeply falling function of the missing energy, so we quote two cases,
\begin{align}\label{eq:SMBR_20MeVcut}
    \mathcal{B}\left(\mufiveeSMfull\, |\, E_{\rm missing}^{\rm true} < 20\text{ MeV}\right) = (8.9 \pm 0.3) \times 10^{-14},
    \\\label{eq:SMBR_10MeVcut}
    \mathcal{B}\left(\mufiveeSMfull\, |\, E_{\rm missing}^{\rm true} < 10\text{ MeV}\right) = (1.1 \pm 0.2) \times 10^{-15},
\end{align}
where $E_{\rm missing}^{\rm true}$ is the true energy of the two neutrinos in the muon rest frame.
As we will see, this rate, even after accounting for detector resolution effects, is far too low and does not constitute a worrisome background to neutrinoless exotic decays.

Finally, we note that muon decays to dark sector particles can also induce $5e2\nu$ final states. 
For example, muon-specific forces \cite{Gninenko:2001hx,Chen:2015vqy,Batell:2016ove,Chen:2017awl} could induce the $\mu^+ \to \overline\nu_\mu \nu_e e^+ S$ decay, where the dark scalar can then decay to four tracks, $S\to 2e^+2e^-$, through a cascade of decays involving other dark particles as intermediate states. 
If the decay/fragmentation of $S$ is fast, such a channel will not be picked up by the proposed missing momentum/missing energy searches \cite{Gninenko:2018tlp,Kahn:2018cqs}. While a more detailed analysis of such decay modes lies outside the scope of the present manuscript, we emphasize that the decay rate, \cref{eq:SM5e}, should be added to the list of physics goals pursued by the Mu3e experiment. For the remainder of the manuscript, we focus on the experimentally cleaner decay mode, the neutrinoless $\mu \to 5e$ decay, which can be induced in dark sector scenarios.

\section{Exotic muon decay: $\mufiveefull$}
\label{sec:exoticdecays}

Previous dark sector searches at muon facilities focused on a single visible or invisible resonance.  
The TWIST and PIENU results provide the best limits on two-body decays to invisible particles, constraining $\mathcal{B}(\mu^+ \to e^+ (X \to {\rm inv})) < \mathcal{O}(10^{-5})$~\cite{TWIST:2014ymv,PIENU:2020loi}.
The MEG experiment searched for $\mu^+ \to e^+ (X \to \gamma \gamma)$, providing the best limits on this branching ratio at the level of $\mathcal{O}(10^{-11})$ for $20 < m_X < 45$~MeV~\cite{MEG:2020zxk}, much above the SM background from double radiative muon decays, $\mathcal{B}(\mu^+ \to \overline{\nu}_\mu \nu_e e^+ \gamma \gamma; E_{\rm missing} < 10 \text{ MeV}) = 1.2\times 10^{-14}$~\cite{Banerjee:2020rww}, where the missing energy $E_{\rm missing}$ is required to be small.
At SINDRUM, constraints on $\mathcal{B}(\mu^+ \to e^+ (X\to e^+e^-)) < \mathcal{O}(10^{-12})$ were placed, depending on the value of $m_X$~\cite{SINDRUM:1986klz}.
The Mu3e experiment will be well positioned to improve on the latter.
Below, we extend this list of phenomenologically  interesting dark sector signatures by considering a cascade decay into the dark sector, $\mu^+ \to e^+ (\hdark \to \Adark \Adark \to 2 (e^+e^-))$.

\subsection{A dark sector model}
\label{sec:model}

Our benchmark model consists of a higgsed dark abelian gauge group $U(1)_d$ coupled to the SM through kinetic mixing between the dark photon and the ordinary photon and through lepton-flavor-violating dimension-five operators.
That is, the Lagrangian of the model is given by $\mathscr{L} = \mathscr{L}_{\rm SM} + \mathscr{L}_{\rm DS} +\mathscr{L}_{\rm LFV}$, where $\mathscr{L}_{\rm SM}$ is the SM Lagrangian and
\begin{equation}\label{eq:DS_lagrangian}
\mathscr{L}_{\rm DS} = (D_\mu\phi)^\dagger D^\mu\phi - \frac{1}{4}F_d^{\mu\nu} F_{d\,\mu\nu}-\frac{\varepsilon}{2} F_d^{\mu\nu} F_{\mu\nu} - \mu^2 \big(\phi^\dagger\phi\big) -\lambda \big(\phi^\dagger\phi\big)^2,
\end{equation}
with $F^{\mu\nu}_d$ the dark photon field strength tensor, while $D^\mu \phi=(\partial^\mu -i g_d  \gamma_{d}'{}^\mu)\phi$, and we set for simplicity the mixed quartic, $\mathscr{L}_{\rm DS} \supset \lambda' \big(\phi^\dagger\phi\big) \big(H^\dagger H\big)$ to zero.
The dark scalar $\phi$ develops a vev, $\phi=(v_d+\hdark)/\sqrt 2$, giving the dark photon a mass $m_{\Adark} = g_d v_d$. 
After the usual field redefinitions \cite{Holdom:1985ag}, we end up with the massive physical dark photon, $\Adark$, the massless photon $\gamma$, and the light scalar $\hdark$.

The LFV interactions are mediated by the dimension-five operators
\begin{equation}\label{eq:LFV_lagrangian}
\mathscr{L}_{\rm LFV} = -\frac{C_{ij}}{\Lambda} \phi \left(\bar{L}_i
H\right) \ell_j  + {\rm h.c.},
\end{equation}
where $\Lambda$ is the cut-off of the effective theory, $L_i$ ($\ell_i$) are the left-handed (right-handed) SM lepton doublets (singlets), and the summation over generation indices $i,j=1,2,3$ is understood. After $\phi$ and $H$ obtain the vevs, the masses of the charged leptons are given by the sum of the SM Yukawas
\beq
{\mathscr L}_{\rm SM}\supset -\lambda_{ij} (\bar L_i H) \ell_j +{\rm h.c.},
\eeq
and the dimension-five terms in \eqref{eq:LFV_lagrangian}, giving
\beq
\text{diag}(m_\ell)=V_L\Big(\lambda+C\frac{v_d}{\sqrt 2 \Lambda}\Big) V_R^\dagger \frac{v}{\sqrt 2},
\eeq
where $V_{L,R}$ are the unitary matrices that diagonalize the mass matrix. Note that the same unitary transformation also diagonalizes the Higgs couplings to leptons.
After mass diagonalization, the interaction Lagrangian in the physical basis is given by
\beq\label{eq:Lint}
{\mathscr L}\supset -m_{\ell_i} \bar \ell_{Li} \ell_{Ri} \Big(1+\frac{h}{v}\Big)- y_{ij} \bar \ell_{Li} \ell_{Rj} h_d\Big(1+\frac{h}{v}\Big) +{\rm h.c.},
\eeq
where $y = V_L C V_R^\dagger {v}/({\sqrt{2}\Lambda})$.

\subsection{Signal rate prediction}

\begin{figure}[t]
    \centering
    \includegraphics[width=0.8\textwidth]{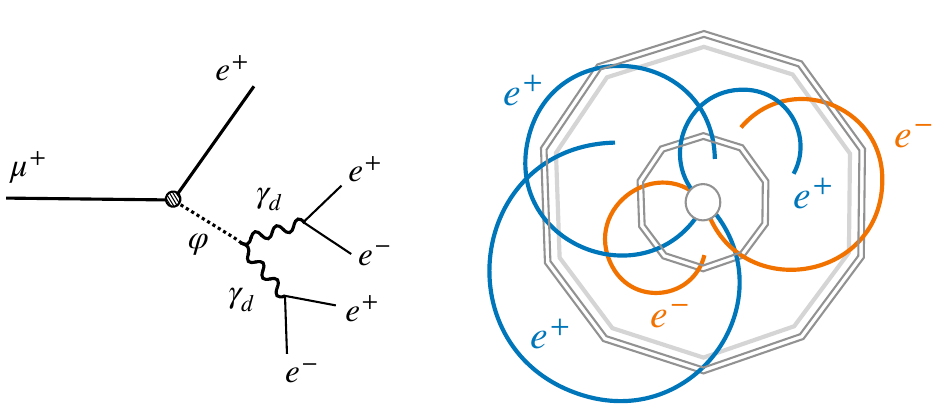}
    \caption{The diagram for the dark scalar production and the subsequent decay chain that realizes $\mu^+ \to e^+ 2(e^+e^-)$ decays (left) and a sketch of the corresponding experimental signature at Mu3e in a transverse view of the detector (not to scale).\label{fig:diagram} }
\end{figure}

The dimension-five flavor-violating couplings \eqref{eq:LFV_lagrangian} in the higgsed dark abelian $U(1)_d$ extension of the SM will lead to the $\ell_i \to \ell_j h_d$ decays if these are kinematically allowed. In this manuscript  we focus on the $\mu \to e h_d$ transition, where  the dark scalar $h_d$ further decays promptly  into two dark photons, each of which decays to an $e^+e^-$ pair, $h_d \to (\Adark \to e^+e^-) (\Adark \to e^+e^-)$, as shown in Fig.~\ref{fig:diagram}. This $\mu \to 5e$ decay can well be the leading signature of such a higgsed dark $U(1)_d$, if the dark gauge coupling is large enough such that the $h_d\to \gamma_d \gamma_d$ decay dominates. The $\mu\to 5e$ cascade decay is kinematically allowed for $m_{h_d} < m_\mu - m_e$ and $m_{h_d} > 2 m_{\Adark}$.

The branching ratio for the $\mu \to e h_d$  decay is given by 
\begin{equation}\label{eq:Br_lepton_decay}
    \mathcal{B}(\mu \to e h_d) = \frac{1}{\Gamma_{\mu}} \,\big( |y_{\mu e}|^2+ |y_{ e\mu}|^2+4 {\rm Re} \big(y_{\mu e}y_{e\mu}^*\big)r_e \big)\frac{m_{\mu}}{32 \pi} \left[(1+r_e)^2 - r_{h_d}^2\right] \lambda^{1/2}(1,r_e^2,r_{h_d}^2),
\end{equation}
where $\Gamma_\mu$ is the muon decay width, $\lambda(x,y,z)$ is the K\"allen function, while $r_{e,h_d} = m_{e,h_d}/m_{\mu}$. Assuming $\mathcal{B}(h_d \to \gamma_d \gamma_d) \approx \mathcal{B}(\gamma_d \to e^+ e^-) \approx 1$, and neglecting the electron mass, the $\mu \to 5e$ branching ratio is then
\begin{equation}\label{eq:Br_muon_decay}
    \mathcal{B}(\mu^+ \to e^+ h_d \to e^+ 2(e^+e^-)) = 3.5 \times 10^{-3} \biggr( \frac{\sqrt{|y_{\mu e}|^2+|y_{ e\mu}|^2}}{ 10^{-9}}\biggr)^2\left(1 - r_{h_d}^2\right)^2.
\end{equation}
As we will discuss, the above numerical example for $y_{\mu e, e\mu}$ is safely below the current experimental bounds discussed in Sec.~\ref{sect:bounds}. 

To a good approximation, the dark scalar decay is prompt as long as the dark coupling is sizeable.
The partial decay width is given by \cite{Batell:2009yf}, 
\begin{equation}\label{eq:Bhtogg}
    \Gamma_{h_d \to 2\gamma_d} = \frac{\alpha_d}{8} \frac{m_{h_d}^3}{m_{\gamma_d}^2}\,f(m_\Adark/m_{h_d}),
\end{equation}
where $f(r) = \left( 1- 4 r^2 + 12 r^4\right) \sqrt{1 - 4 r^2}$.
In the limit of $h_d\to \gamma_d\gamma_d$ dominating the decay width, this then also gives the lifetime of the dark Higgs,
\begin{equation}
    c\tau_{h_d} = 270 \text{ fm}\,\times \left(\frac{\alpha}{\alpha_D}\right)
     \left(\frac{90\text{ MeV}}{m_{h_d}}\right) 
    \left(\frac{3}{m_{h_d}/m_\Adark}\right)^2 
    \frac{1}{f(m_\Adark/m_{h_d})}.
\end{equation}
On the other hand, the dark photon decays back to the SM particles through the kinetic mixing parameter, $\varepsilon \lesssim 10^{-3}$, and, therefore, may have an observable displacement within the experiment.
In the mass range of interest, $ 2 m_e < m_\Adark \lesssim (m_\mu- m_e)/2$, the dark photon always decays back to electron-positron pairs, $\gamma_d \to e^+e^-$, with the decay width \cite{Batell:2009yf}
\begin{equation}
    \Gamma_{\gamma_d \to e^+e^-} = \frac{\alpha \varepsilon^2 m_{\gamma_d}}{3} \left( 1 - 4 r^2_e \right)^{1/2} \left( 1 + 2 r_e^2 \right),
\end{equation}
giving the $\gamma_d$ lifetime of
\begin{equation}
c \tau_\Adark^0 = 0.27 \,\, \text{mm}\,\times \left(\frac{10^{-4}}{\varepsilon}\right)^2 \left( \frac{30 \text{ MeV}}{m_\Adark}\right)\frac{1}{g(m_{e}/m_\Adark)},
\end{equation}
where $g(r) = \left(1 + 2r^2\right) \sqrt{ 1 - 4r^2}$.
Most dark photons would still decay within the stopping target for the above parameters, leading to no significant modification of the $\mufivee$ signals.
For smaller kinetic mixing parameters $\varepsilon \ll 10^{-4}$, other constraints from beam-dump and fixed-target experiments exclude the mass range of interest~\cite{Petersen:2023hgm}.
Therefore, in our discussion, we will always assume that the dark photons decay promptly inside the Mu3e target.

\subsection{Other experimental bounds}\label{sect:bounds}
The $y_{\mu e,e\mu}$ LFV couplings of the dark Higgs are constrained by a few other probes, with the most relevant constraints reviewed below. 
However, these bounds are not very stringent and do not significantly limit the possible strength of the $\mu \to 5e$ signal. 

\paragraph{Muon lifetime:} 
Simply requiring that the $\mu\to 5e$ decay does not saturate the muon decay width ($\Gamma_{\mu \to e \hdark} < \Gamma_\mu$), gives $\sqrt{|y_{\mu e}|^2+|y_{ e\mu}|^2}<1.7 \times 10^{-8}$, which by itself is already quite constraining. 
Note that a factor of ${\mathcal O}(30)$ tighter bound on $(|y_{\mu e}|^2+|y_{ e\mu}|^2)^{1/2}$ can be extracted from the consistency of the Fermi constant extracted from the muon lifetime measurement confronted with other determinations of $G_F$ \cite{Efrati:2015eaa,Balkin:2022glu}.

\paragraph{Neutrinoless muon decays:}
An experiment at Lawrence Berkeley Laboratory performed several searches for rare muon decays, including the channel $\mu^+ \to e^+ \gamma \gamma $~\cite{Poutissou:1974ic}.
The detector constituted a cylindrical NaI crystal that surrounded the muon target.
The constraint was based on an inclusive requirement that the entire energy collected in the NaI crystal reconstruct the muon mass.
This results in a limit on neutrinoless muon decays to any number of electrons and photons, from which we conclude that $\mathcal{B}(\mufivee) < 4 \times 10^{-6}$.
Other searches for the two-photon mode were performed at an experiment in TRIUMF~\cite{Azuelos:1983wx}, at the Crystal Box detector at Los Alamos~\cite{Bolton:1988af}, and more recently at the MEG~\cite{MEG:2020zxk} experiment at PSI.
Neither of these limits directly applies to the $5e$ mode due to the ability of the detectors to differentiate between charged and neutral particles.
A similar argument applies to recasts of $\mu \to e \gamma$ searches~\cite{Bowman:1978kz}.

In principle, the annihilation of muonium $\mu^+ e^- \to \Adark \Adark \to 2e^+ 2e^-$ can also provide a limit on our model, though the rate is suppressed by the wave function overlap. 
We are not aware of any searches for this channel. 
The analogous $\mu^+e^- \to \gamma \gamma$ decay has been constrained in Ref.~\cite{PhysRevLett.3.288}, but their limit does not directly apply to the final state with charged particles.

\paragraph{Higgs decays:} The Higgs Yukawas are diagonal in the mass basis, \eqref{eq:Lint}, so that the $h\to \mu e$ decays are forbidden. Flavor-violating Higgs decays are possible, if they are accompanied by an emission of a dark Higgs, $h\to \mu e h_d$. Such three-body decays will smear the $\mu e$ invariant mass distribution and thus a reduced signal in the $h\to \mu e$ searches at the LHC. While the full recasting of the bounds is beyond the scope of this paper, we can get a rough estimate of the exclusion by assuming that the effect is similar in size to what was found for the $h\to \tau \mu \phi$ decays, with $\phi$ an invisible particle, in Ref. \cite{Galon:2017qes}. Under this approximation, the bound ${\cal B}(h\to \mu e)<4.4\times 10^{-5}$ \cite{CMS:2023pte} (see also~\cite{ATLAS:2019old}) would imply $\big(|y_{\mu e}|^2+|y_{e\mu}|^2\big)^{1/2}\lesssim 0.03$.

In principle, there is also sensitivity to $h_d -\mu\mu$ coupling, $y_{\mu\mu}$, from the measured $h\to \mu\mu$ branching ratio, $\mathcal{B}(h \to \mu \mu) = (2.6\pm 1.3) \times 10^{-4}$ \cite{ATLAS:2022vkf}. Some of the $h\to \mu\mu h_d$ bremsstrahlung decays would pass experimental cuts and be part of the $h\to \mu\mu$ signal. 
Taking as a rough estimate that the efficiency of the experimental cuts is similar as for $h\to \tau \mu \phi$~\cite{Galon:2017qes}, this would then lead to a very weak bound, $y_{\mu\mu}\lesssim {\mathcal O}(0.1)$.

\paragraph{$\boldsymbol{\mu \rightarrow e \gamma}$:} The LFV Yukawa couplings of $h_d$ in \cref{eq:Lint} mediate $\mu\to e\gamma$ decays at the one-loop level via the effective Lagrangian~\cite{Harnik:2012pb}
\begin{equation}
    \mathcal{L}_\gamma = \frac{e m_\mu}{8\pi^2} (c_L \bar{e} \sigma^{\alpha\beta} P_L \mu + c_R \bar{e} \sigma^{\alpha\beta} P_R \mu)F_{\alpha \beta} + \text{h.c.},
\end{equation}
where $c_{L,R}$ are dimensionful Wilson coefficients. The rate for $\mu \to e \gamma$ is given by 
\begin{equation}
    \Gamma_{\mu \to e \gamma} = \frac{\alpha m_\mu^5}{64\pi^4} (|c_L|^2 + |c_R|^2),
\end{equation}
where, assuming $y_{\mu\mu} \gg y_{ee}$ for simplicity, and $m_e \ll m_\hdark \ll m_\mu$ the Wilson coefficient $c_L$ can be expressed as
\begin{equation}
    c_L \approx \frac{1}{8 m_\mu^2} y^*_{\mu e}\Big[ \text{Re} (y_{\mu\mu}) \left( - 5 + \frac{2\pi^2}{3} \right)- i \text{Im} (y_{\mu\mu})\Big]
\end{equation}
and $c_R$ given equivalently with the substitution $y^*_{\mu e} \rightarrow y_{e\mu}$. Using the final limits from MEG~\cite{MEG:2016leq} constraining $\mathcal{B}(\mu\to e\gamma) < 4.2 \times 10^{-13}$ and assuming $\text{Im}(y_{\mu\mu}) = 0$, we obtain
\begin{equation}
    y_{\mu\mu}\sqrt{|y_{\mu e}|^2 + |y_{e\mu}|^2} < 5.1 \times 10^{-12}.
\end{equation}

\paragraph{$\boldsymbol{\mu \rightarrow 3e}$:} The decay $\mu \to e h_d$, followed by $h_d\to e^+e^-$ decay, gives a constraint on a combination of $y_{\mu e}$ and $y_{ee}$ couplings. We are interested in the regime, where $h_d\to \gamma_d\gamma_d$ dominates so that only a small fraction of $h_d$ decays through the $h_d\to e^+e^-$ channel.  
The $h_d\to e^+e^-$ branching ratio, in the regime where $\mathcal{B}(h_d \to \gamma_d \gamma_d) \approx 1$, is
\begin{equation}
 \mathcal{B}( h_d\to e^+e^-)\simeq \frac{ 4|y_{ee}|^2 m_{\Adark}^2}{g_d^2 m_{\hdark}^2}\frac{1}{f(m_\Adark/m_{h_d})},
\end{equation}
where for simplicity we took the limit $m_e \ll m_{h_d}$.
The bound $\mathcal{B}(\mu \to 3e) <1.0 \times 10^{-12} $ then translates to
\beq
\big( |y_{\mu e}|^2+ |y_{ e\mu}|^2\big)^{1/2}< 5\times 10^{-6} \times \biggr( \frac{10^{-9}}{y_{ee}}\biggr) \biggr( \frac{g_d}{0.1}\biggr) \biggr( \frac{m_{\gamma_d}}{5\, \text{MeV}}\biggr)\biggr( \frac{30\, \text{MeV}}{m_{h_d}}\biggr).
\eeq
Note that this simple scaling of the bound is valid for $m_{\gamma_d}\sim {\mathcal O}(m_{h_d})$ as long at $y_{ee}\ll g_d$.

\section{Experimental Reach}
\label{sec:experimentalreach}

In this section, we address the experimental reach of Mu3e to the five-track channels.
We start with a general discussion of the expected backgrounds to $\mufiveeSM$ and $\mufivee$ measurements at Mu3e in \cref{sec:backgrounds}.
We leave a more detailed background rate estimate for future studies within the Mu3e collaboration.
In the remainder of the section, we present our assumptions behind our sensitivity estimates.
\Cref{sec:simulation} describes our simplified numerical simulation of Mu3e.
We then present the selection criterion and resulting efficiencies for both SM and new-physics decays in \cref{sec:sensitivity}.
We then present a few future directions that can improve upon our sensitivity estimates in \cref{sec:future}.

\subsection{Experimental backgrounds}
\label{sec:backgrounds}

Experimental backgrounds for the $\mufivee$ process can be split into two categories: 1. irreducible backgrounds from radiative muon decays where two photons internally convert into two electron-positron pairs and 2. accidental combinatorial backgrounds from, for example, proximal radiative and Michel decays, proximal Michel decays in combination with Bhabha scatterings, or Michel decays with detector reconstruction deficiencies such as charge misidentification.

\paragraph{Decays with missing energy:}
Backgrounds from $\mufiveeSM$ may be mitigated by exploring the different kinematics of the neutrinoless $\mufivee$ decay.
Of greatest significance is the cut on the reconstructed missing energy, or equivalently, on the reconstructed energy of the five-electron system.
The left panel in \cref{fig:missing_E_dists} shows the missing energy distribution of the signal and the background.
The signal events are compatible with no missing energy within the momentum resolution of the detector.
Using our simulation, we find that a cut on the missing energy of $10$~MeV can suppress the $\mufiveeSM$ background down to the $\mathcal{O}(10^{-15})$ level.
This is compatible with the branching ratios in \cref{eq:SMBR_20MeVcut,eq:SMBR_10MeVcut}.

Due to the significant missing energy carried by neutrinos, most electron and positron tracks are too soft to be reconstructed as short or long tracks --- they do not travel far enough in the transverse direction.
As shown in the right panel in \cref{fig:missing_E_dists}, this is particularly true for the electron tracks, which are a product of internal photon conversion.
While this hurts the sensitivity to the SM decays, it becomes an advantage in searches for the exotic decays.

\paragraph{Accidental backgrounds:}
Interactions of Michel decay positrons with the target and detector material are an important  
source of backgrounds to the $\mu \to 3e$ signal~\cite{Mu3e:2020gyw}, and therefore should be carefully estimated for the $\mu\to 5e$ channel.
For the $3e$ channel, it was shown by the Mu3e collaboration that this background  can be fully mitigated, {\em i.e.}, reduced to below 0.2 events per $10^{15}$ muon decays using kinematics, timing, and position cuts.
For the $5e$ channel, such cuts are also expected to be important since the momenta of the five tracks also need to sum up to zero.
The $5e$ channel, however, has an advantage compared to the $3e$ one: the $5e$ event contains two negatively charged tracks in the event, which is much less common in the experiment.
To take advantage of this feature, a good understanding of the charge misidentification capabilities of Mu3e will be crucial.
However, even if/when positrons get misreconstructed as electrons, the kinematic, positional, and timing cuts remain available.

In what follows, we provide a naive and conservative estimate of the total number of accidental backgrounds per $10^{15}$ muon decays, following a similar prescription to the one outlined in Ref.~\cite{Echenard:2014lma}.
We neglect Compton scattering and photon conversion within the target and detector material, which are subdominant compared to Bhabha scattering. 
The possible sources of accidental backgrounds are 
\begin{enumerate}
    \item A $\mu \to 3e 2\nu$ decay in combination with a proximal Michel decay, where the ejected positron immediately Bhabha scatters on an electron within the target material,
    \begin{equation}
        N_{\mu \to 3e 2\nu+\text{Bhabha}} = N_\mu  \times (R_\mu \delta t) \times P_p \times ( {\mathcal B}_{\mu\to 3e 2\nu} {\mathcal B}_{\mu \rightarrow 1e2\nu}) \times P_{\rm Bhabha} \approx 10^{3}.
    \end{equation}
    Alternatively, one could also consider a single Michel decay in combination with a $\mu \to 3e 2\nu$ decay, where the Bhabha scattering is initiated by one of the two positrons produced in $\mu \to 3e 2\nu$.
    This brings the total rate to approximately $3\times 10^{3}$ events, before any kinematical cuts, and assuming that the positrons from $\mu \to 3e 2\nu$ can produce visible $e^+e^-$ Bhabha pairs as often as the Michel positrons.
    \item Three Michel decays where two of the three positrons have Bhabha scattered with electrons within the target material,
    \begin{equation}
        N_{3M+\text{Bhabha}} = \frac{1}{2} N_\mu \times (R_\mu \delta t)^2 \times P_p^2 \times ({\mathcal B}_{\mu \rightarrow 1e 2\nu})^3 \times  P_{\rm Bhabha}^2 \approx 10^{-1}.
    \end{equation}
    \item A Michel decay in combination with two radiative Michel decays, where the two photons convert into electron-positron pairs within the target, and the two positrons 
    from these conversions remain undetected,
    \begin{equation}
        N_{3M_\gamma} = \frac{2}{3} N_\mu \times (R_\mu \delta t)^2 \times P_p^2 \times ({\mathcal B}_{\mu \rightarrow 1e 2\nu} {\mathcal B}^2_{\mu\rightarrow 1e 2\nu 1\gamma}) \times  P_\gamma^2 \approx 10^{-3}.
    \end{equation}
    \item Five Michel decays with two positrons misidentified as electrons,
    \begin{equation}
        N_{5M} = \frac{1}{12} N_\mu \times (R_\mu \delta t)^4 \times P_p^4  \times ({\mathcal B}_{\mu \rightarrow 1e 2\nu} ) ^5 \times P^2_{e^+ \rightarrow e^-} \approx 10^{-4}.
    \end{equation}
    \end{enumerate}
Above, the total number of muons was assumed to be $N_\mu = 10^{15}$, with $R_\mu = 10^8$/s the stopped muon rate, $\delta t = 2.5 \times 10^{-10}$~s the average time resolution, $P_p = 10^{-2}$ the vertex timing resolution suppression factor (taken as the quoted suppression power of the fiber detectors~\cite{Mu3e:2020gyw}), $P_{e^+ \rightarrow e^-} = 0.45 \%$ the charge misidentification probability, ${\mathcal B}_{\mu \rightarrow 1e 2\nu}\approx 1$, ${\mathcal B}_{\mu \rightarrow 1e 2\nu \gamma}\approx 0.014$ ($E_\gamma > 10$ MeV) \cite{ParticleDataGroup:2022pth}, $P_{\rm Bhabha} \approx 10^{-4}$ is the conditional probability for an observable Bhabha scattered $e^+ e^-$ pair given a positron from within the target (from Tab. 22.2 of Ref.~\cite{Mu3e:2020gyw}), while $P_\gamma = 8 \times 10^{-4}$ is the assumed photon conversion probability within the target. 
Note that the above estimates hold in the limit $R_\mu T \gg 1$ where $T$ is the duration of the measurement (roughly $\pi \times 10^7$s for phase I of Mu3e to which also the above choices of parameters apply).

In addition to the coincident combination of Bhabha processes and radiative modes of muon decays, two electrons can also be observed when an electron produced in $\mu\ \to 3e 2\nu$ decays undergoes M\o{}ller scattering or initiates trident production in the material. 
Because the number of high-momentum electrons in the experiment is far smaller than that of positrons, we expect these channels to be subdominant to the background sources listed above.

In their $3e$ sensitivity estimates, the Mu3e collaboration also included a cut on the lowest invariant pairwise $e^+ e^-$ mass in order to suppress Bhabha pair production backgrounds in the window from $5$~MeV to $10$~MeV.
This will limit the sensitivity to new physics for dark photon masses within this range if such a cut is still necessary for the $5e$ channel.
While a full detector simulation, including material effects and vertex reconstruction, is needed in order to accurately assess the loss in sensitivity due to these backgrounds, we do expect the double coincidence in $e^+e^-$ invariant masses to be a sufficiently strong discriminator such that one could efficiently search for new-physics events.

\subsection{Mu3e simulation}
\label{sec:simulation}

\begin{figure}[t]
    \centering
    \includegraphics[width=.95\textwidth]{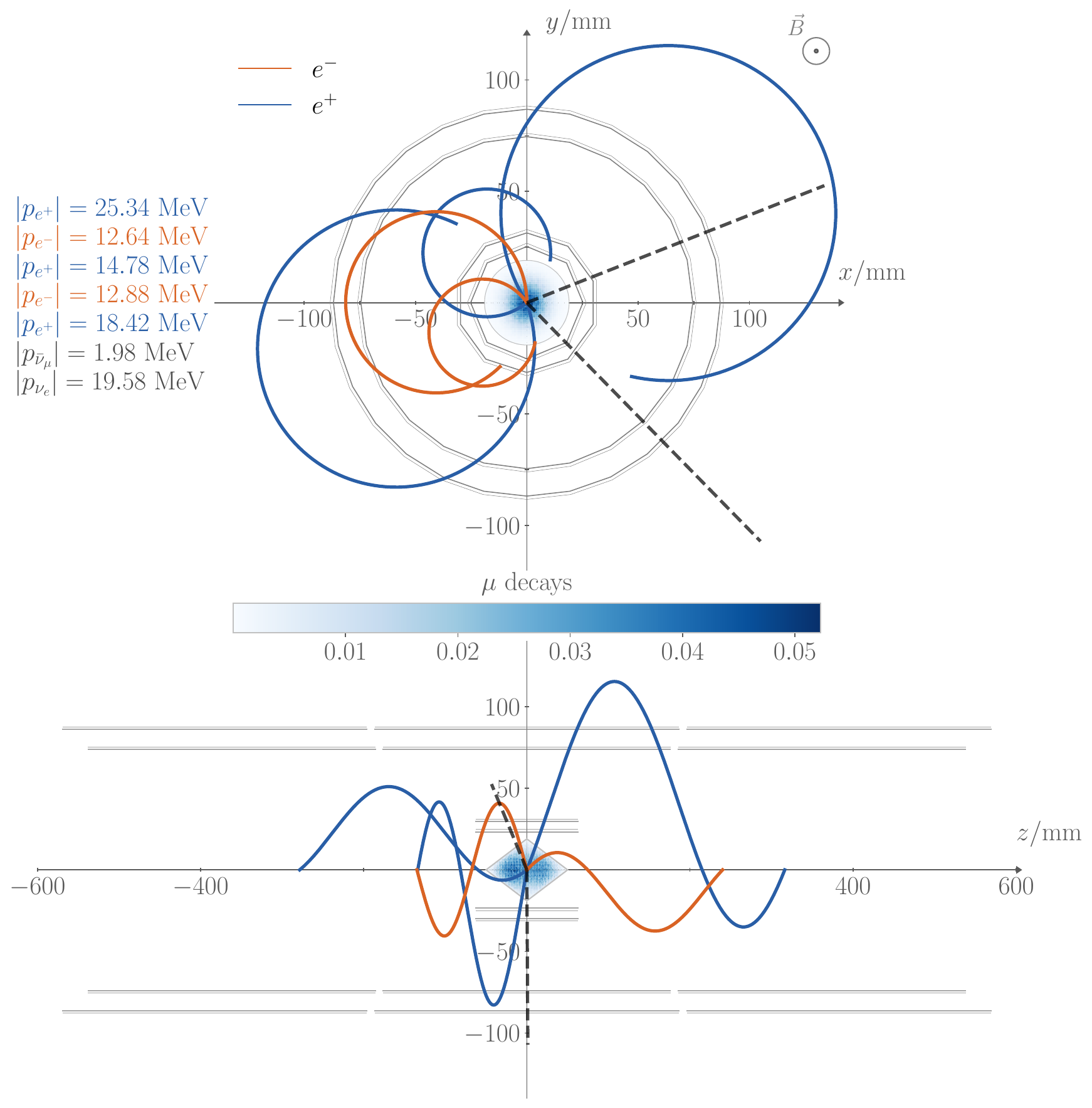}
    \caption{Transverse (top) and longitudinal (bottom) event displays showing an example of a simulated SM five-track $\mu^+\to 3e^+2e^-2\nu$ decay event at Mu3e. The blue heatmap within the double-cone target of each panel represents the modeled $x,y,$ and $z$ probability densities of muon decays within the double-cone target.
    \label{fig:event_displays}
    }
\end{figure}

To study the sensitivity of Mu3e to muon decays with five tracks, we built a fast MC of the detector and simulate $\mu^+$ decays using \textsc{MadGraph5\_aMC@NLO} for $\mufiveeSM$ and the \textsc{phasespace}~\cite{puig_eschle_phasespace-2019} package for $\muthreee$.
We simulate a total of $13$ million events for the former and $1$ million for the latter.
The simulation distributes muons inside the target, modeled as a hollow double-cone structure.
The target's maximum radius is $19$~mm, and its total length is $100$~mm.
The $z$ position of the decays inside the target is sampled according to Fig.~6.3 of Ref.~\cite{Mu3e:2020gyw} that shows the simulated stopping distribution along the $z$-direction of the double-cone target.
The $x$ and $y$ positions are sampled according to the transverse beam profile modeled as a two-dimensional Gaussian with widths $\sigma_x = 7.50$~mm and $\sigma_y = 8.74$~mm.
The probability density of muon decays inside the target is shown as a heatmap in the event displays in \cref{fig:event_displays}.

Each electron and positron trajectory with the initial position $(x_0, y_0, z_0)$ and momentum $(p_x, p_y, p_z)$ is determined by the equations of motion
\begin{align}
    x(t) &= x_0 + \frac{p_{y}}{B q} \left\{ 1 - \cos\left( \frac{B q}{m} t \right) + \frac{p_{x}}{p_{y}} \sin\left( \frac{B q}{m} t \right) \right\}, 
    \\
    y(t) &= y_0 + \frac{p_{x}}{B q} \left\{ -1 + \cos\left( \frac{B q}{m} t \right) + \frac{p_{y}}{p_{x}} \sin\left( \frac{B q}{m} t \right) \right\}
    \\
    z(t) &= z_0 + \frac{p_z}{m} t,
\end{align}
where $B$ is the magnitude of the magnetic field ($B = 1.0$~T) directed along the beamline, $q$ is the electric charge, and $m$ is the particle mass. The particles follow a helical trajectory with radius $R = p_T/B$, where $p_T \equiv ((p_x)^2 + (p_y)^2)^{1/2}$ is the total momentum in the $x-y$ plane, transverse to the beam pipe.
The trajectory length is determined by the point at which the particle exits the detector barrels in the $z$ direction, defined as the end of the longest recurler cylinder.
We neglect any particle losses inside the barrel.
Having determined the trajectory in the $x-y$ plane, we count the number of intersections between each track and the active components of the detector, each defining a hit. 
The primary active layers are modeled as four thin cylinders centered around the double-cone target.
From the innermost to the outermost layers, the cylinders have radii of \radiuslayerone, \radiuslayertwo, \radiuslayerthree, and \radiuslayerfour, and extents of \lengthlayerone, \lengthlayertwo, \lengthlayerthree, and \lengthlayerfour.
In addition to these layers, Mu3e plans to install recurler components, extending the outermost cylinders up and downstream of the target, thereby increasing the probability of detecting tracks that curl back at large values of $z$.
We replicate the two outermost cylinders with a shifted $z$ location to account for these. 
The gap between the end of the outermost detector and the recurler is assumed to be $20$~mm.

The momentum resolution of the tracks depends on the number of hits for a given trajectory. 
Tracks with at least 4 hits constitute a short track, and the resolution is $\sigma_p/p \sim 5\%$.
We smear them according to the resolution in Fig.~19.2 of Ref.~\cite{Mu3e:2020gyw}.
The resolution can be far better for long tracks, defined by tracks with 6 and 8 hits, ranging from $\sigma_p/p \sim 0.5\%$ to $\sigma_p/p \sim 3\%$.
We smear these tracks following Fig.~19.3 of Ref.~\cite{Mu3e:2020gyw}.
Our analysis assumes that the energy losses are fully accounted for and that the detection time is not used.
We have checked that our setup reproduces the correct reconstruction gaps in the $\cos{\theta}$ and $p_e$ plane, in accordance with Fig.~19.4 of Ref.~\cite{Mu3e:2020gyw}.

\begin{figure}[t]
    \centering
    \includegraphics[width=0.49\textwidth]{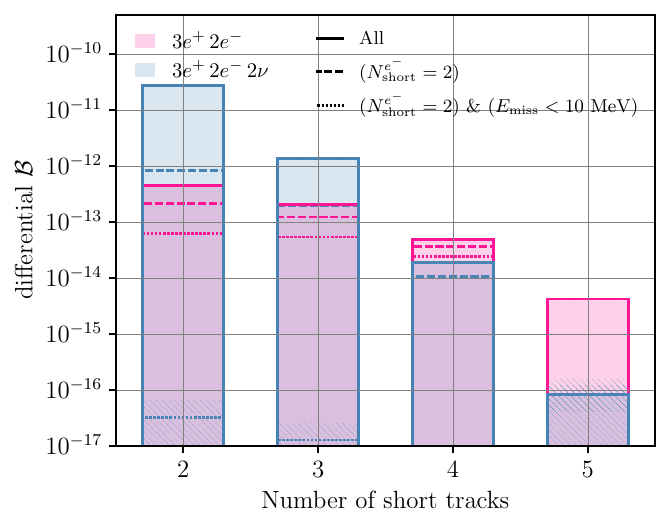}
    \includegraphics[width=0.49\textwidth]{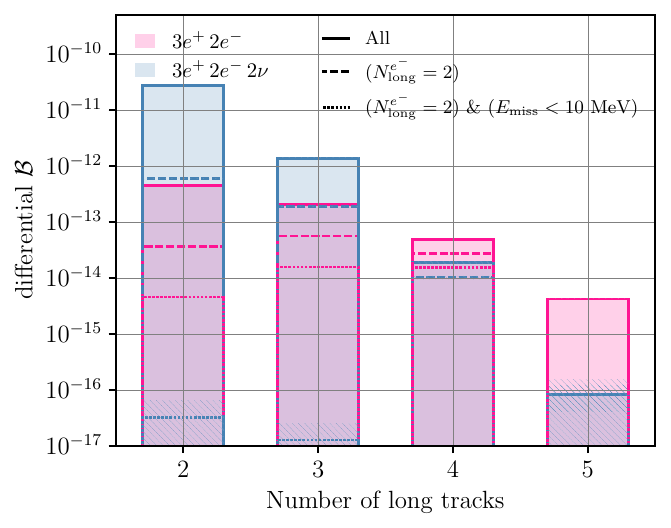}
    \caption{
    The number of electrons or positrons that are reconstructed as short tracks (left) and long tracks (right) for $\mufivee$ (pink) and $\mufiveeSM$ (blue) decays. 
    We show the new-physics decays for $m_\Adark = 30$~MeV and $m_\hdark = 90$~MeV.
    A short track is defined by $N_{\rm hits} \geq 4$ and a long track as $N_{\rm hits} \geq 6$.
    Solid lines show the total branching ratio for short or long tracks.
    Dashed lines have an additional requirement that at least two \emph{electrons} be reconstructed as either short or long tracks.
    Dotted lines have yet another requirement on the missing energy $E_{\rm miss}$, defined as the difference between the total energy of short tracks and the muon mass.
    The hatched bands indicate the MC statistical uncertainty.
    \label{fig:track_dist}}
\end{figure}

\begin{figure}[t]
    \centering
    \includegraphics[width=0.49\textwidth]{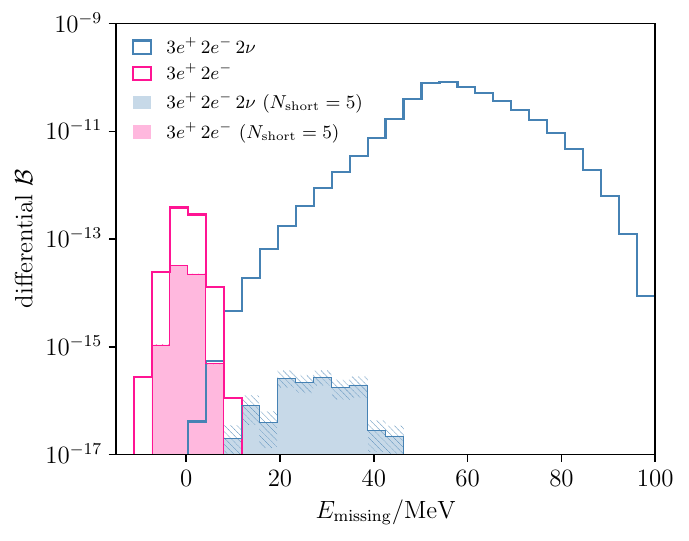}
    \includegraphics[width=0.49\textwidth]{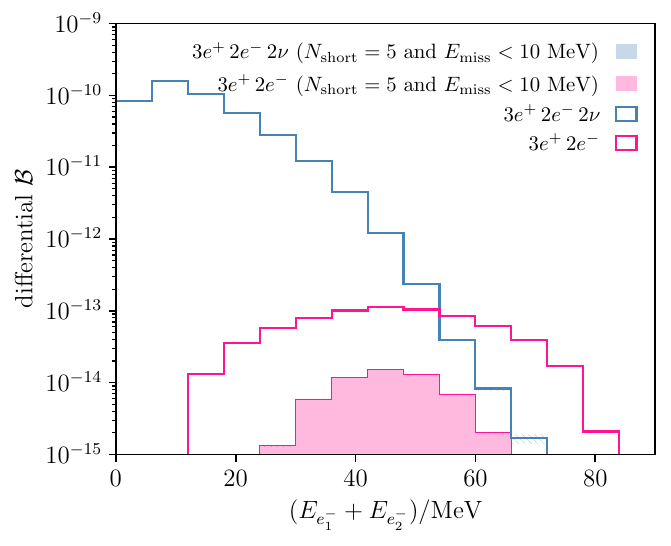}
    \caption{Left panel shows the reconstructed missing energy distribution for $\mufivee$ ($m_\Adark = 30$~MeV and $m_\hdark = 90$~MeV) compared with the $\mu^+\to 3e^+ 2e^- 2\nu$ SM rate. 
    Solid histograms include the requirement of exactly 5 reconstructed short tracks.
    In the right panel we show the reconstructed total electron energy distribution.
   In this case, solid histograms require exactly 5 reconstructed short-tracks and $E_{\rm missing} < 10$~MeV.
    \label{fig:missing_E_dists} 
    }
\end{figure}

\subsection{The Mu3e sensitivity}
\label{sec:sensitivity}

\paragraph{Sensitivity to SM decays:} \label{par:SMsensitivity}
While we do not use a full background simulation, we can still make a few low-level requirements that suppress intrinsic muon-decay backgrounds to the exotic $\mufivee$ decays.
For the SM rate $\mufiveeSM$, we require that all five tracks are reconstructed as short tracks ($n_{\rm hits} \geq 4$ for each track).
In other words, all tracks are required to cross at least four layers of the detector.
This automatically imposes a momentum threshold of about $10$~MeV for each charged track and already significantly reduces background with missing energy as shown in \cref{fig:track_dist}.
Our simulation shows that the resulting branching ratio for five short-track events from the process $\mufiveeSMfull$ is about $\mathcal{B}\sim 2 \times 10^{-15}$, much below the value in \cref{eq:SMBR_five_tracks}, which uses truth-level cuts.
This loss in rate is due to the steep dependence of the rate on the transverse-momentum threshold and the muon decay position within the target.
Whether this five-track rate is detectable or not will rely on the resulting efficiencies of the five-track reconstruction algorithm and the accidental and beam-induced backgrounds.
We also note that even when all tracks have sufficient energy to fall within the detector acceptance, the amount of missing energy in these events is still significant, as shown in \cref{fig:missing_E_dists}.
We comment on potential alternative directions to detect the SM rate in \cref{sec:future}.

\begin{figure}[t]
    \centering
    \includegraphics[width=\textwidth]{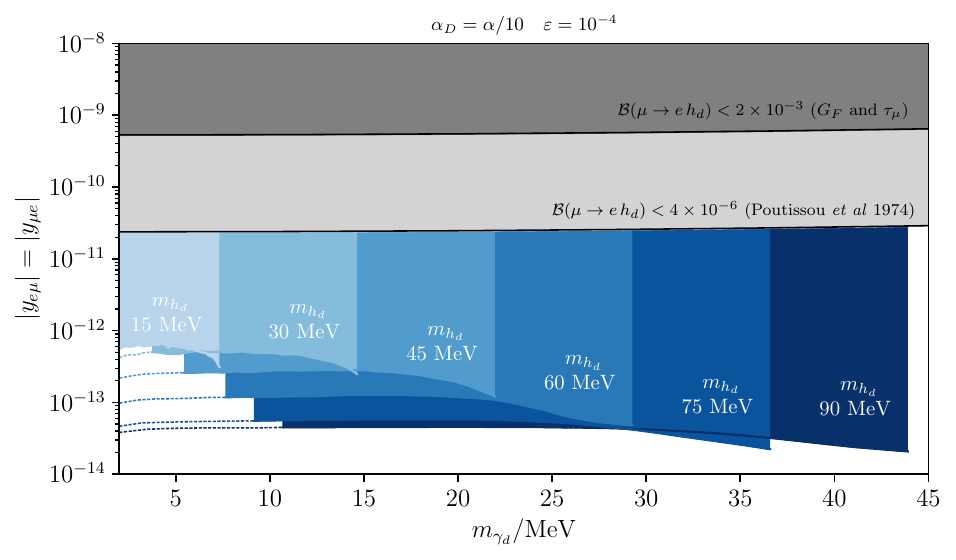}
    \caption{The Mu3e reach for $\mu^+ \to e^+ (\hdark \to \Adark \Adark)$ decays as a function of $m_\Adark$ and $y_{\mu e}=y_{e\mu}$ for various choices of $m_\hdark$, assuming a sensitivity of $\mathcal{B}(\mu^+ \to e^+ h_d) < 10^{-12}$ after signal selection described in \cref{sec:simulation}. 
    We fix $\varepsilon = 10^{-4}$ and $\alpha_D = \alpha / 10$.
    Dashed lines indicate regions where the scalar quartic coupling $\lambda > \sqrt{4 \pi}$ for this particular choice of $\alpha_D$. 
    The limit based on a comparison of the total muon lifetime $\tau_\mu$ and independent determinations of $G_F$ is shown as a dark grey region ($\mathcal{B} < 2\times 10^{-3}$) and the limit from Ref.~\cite{Poutissou:1974ic} is shown in the light grey region ($\mathcal{B} < 4\times 10^{-6}$).
    \label{fig:sensitivity}}
\end{figure}

\paragraph{Sensitivity to exotic decays:} \label{par:newphysics}
Similarly to the SM decays above, to estimate the sensitivity of Mu3e to the neutrinoless channel, we impose the requirement that all five tracks are within the experimental acceptance. 
Since the exotic mode has no missing energy, we further require that $E_{\rm missing} < 10$~MeV, similarly to the sensitivity studies for $\muthreee$.
The impact of this last cut on the new-physics rate as well as on the intrinsic SM background is shown in \cref{fig:missing_E_dists}.
Even though the MC statistics are small in this region of phase space, we can readily conclude that the combination of applied cuts is exceedingly stringent, pushing the estimated intrinsic background branching ratio to $\mathcal{O}(10^{-15})$ or lower at no significance cost to the signal efficiency.

We quote the sensitivity of Mu3e to lepton-flavor-violating decays with five tracks requiring a total of $10^{3}$ new-physics events per $10^{15}$ $\mu^+$ decays after the signal selection discussed above.
We believe that this requirement is extremely conservative.
Firstly, the signal selection criteria can be relaxed by exploiting charge identification (most new-physics events have two visible electron tracks), significantly boosting the signal efficiency.
Secondly, while we do not simulate accidental, beam-induced, and charge-misidentification backgrounds, our discussion in \cref{sec:backgrounds} indicates that even before any kinematical requirements, these backgrounds are not expected to be larger than about a thousand events.
Therefore, if the power of kinematical background rejection that was found for the $3e$ channel is of similar size also for the $5e$ channel, then the $\mu \to 5e$ search should be background free for branching ratios as low as $10^{-15}$.
Nevertheless, all our results are based on a signal rate of 1000 events, post-selection.

\begin{figure}[t]
    \centering
    \includegraphics[width=0.49\textwidth]{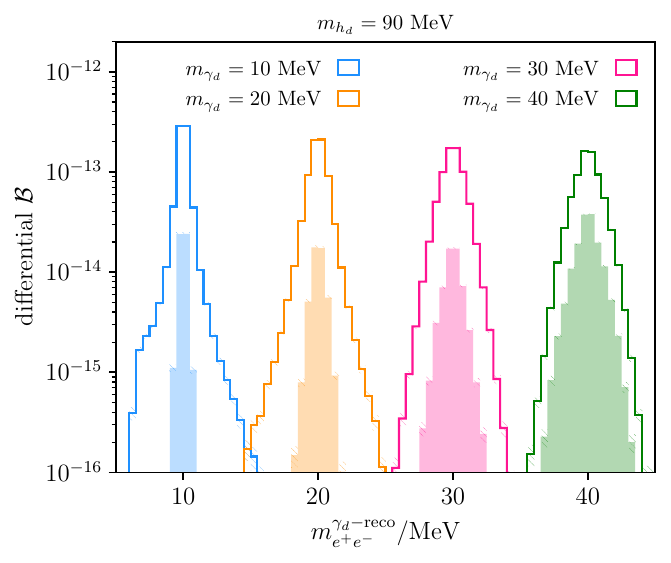}
    \includegraphics[width=0.49\textwidth]{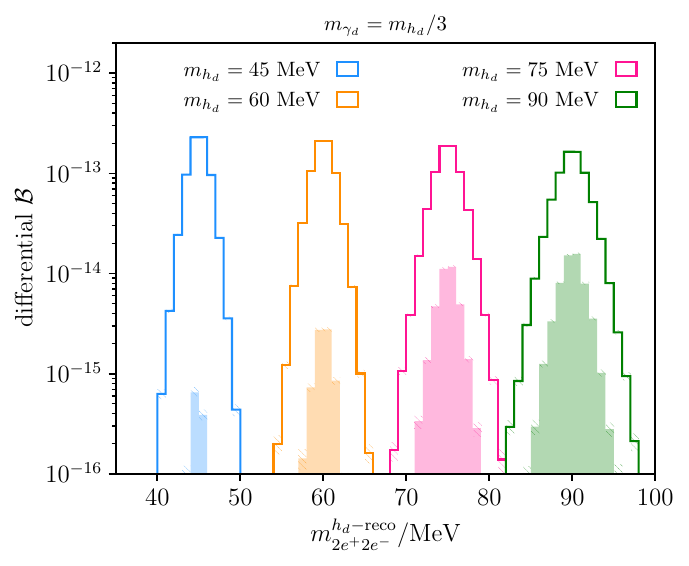}
    \caption{The reconstructed masses of the dark photon (left) and the dark higgs (right) for various fixed masses before signal selection cuts (solid lines) and after signal selection cuts (filled histograms).
    After selection, the resulting resolution on the dark photon mass 
    is approximately $\sigma_{m_{\gamma_d}}/m_{\gamma_d} = 2.3\%$ and $\sigma_{m_{h_d}}/m_{h_d} = 1.5\%$ for the dark higgs mass. 
    The total branching ratio for the signal was fixed to $7\times 10^{-13}$ in both panels.
    \label{fig:mass_resolution}}
\end{figure}

The reach of Mu3e in the parameter space of the dark higgsed $U(1)_d$ model, discussed in \cref{sec:model}, is shown in \cref{fig:sensitivity} for different values of $m_\hdark$ and $m_\Adark$.
We fix the kinetic mixing parameter to $\varepsilon =10^{-4}$, below the reach of the prompt decay searches, and above the limits set by the beam-dump searches at  FASER~\cite{CERN-FASER-CONF-2023-001} and other beam-dump experiments~\cite{Dobrich:2023dkm,Ilten:2018crw,Antel:2023hkf}.
We also fix the dark coupling to $\alpha_D = \alpha/10$ for illustration, although the decay rate is mostly insensitive to the exact choice of this parameter. 
In the regions of parameter space where this choice leads to a breakdown of perturbativity for the scalar quartic coupling, $\lambda > \sqrt{4 \pi}$, we draw the experimental sensitivity with a dashed line. No previous experimental search for this channel exists to the best of our knowledge, so Mu3e can provide the best limits on $|y_{e\mu}|^2 + |y_{\mu e}|^2$ for $m_{h_d} < m_{\mu} - m_e$ and $m_\Adark <m_\hdark /2$.
In the case of a discovery, Mu3e could measure the dark photon and dark scalar masses, depending on the observed signal rate.
We show the reconstructed dark photon and scalar masses in \cref{fig:mass_resolution}.

\subsection{Future directions}
\label{sec:future}

Finally, we also comment on the possibility that searching for exotic multi-lepton production need not require five tracks.
Depending on the charge misidentification capabilities, two electron tracks should already constitute a sufficiently-exotic final state.
In that case, the selection criterion of requiring exactly five tracks in the event may be relaxed to requiring \emph{at most} five tracks but exactly two electrons.
For instance, with this more flexible selection, reconstructing an $e^+e^-e^+e^-$ event with a missed positron as missing energy may be sufficient.
With the alternative strategy to require two negatively charged tracks and relax the requirement on the total number of tracks, the observable branching ratios can be much larger.
Requiring only two positrons and two electrons boosts the observable branching ratio to $7 \times 10^{-14}$.
Similarly, requiring only one positron and two electrons recovers another order of magnitude, with a branching ratio of $8\times 10^{-13}$.

\begin{figure}[t]
    \centering
    \includegraphics[width=0.6\textwidth]{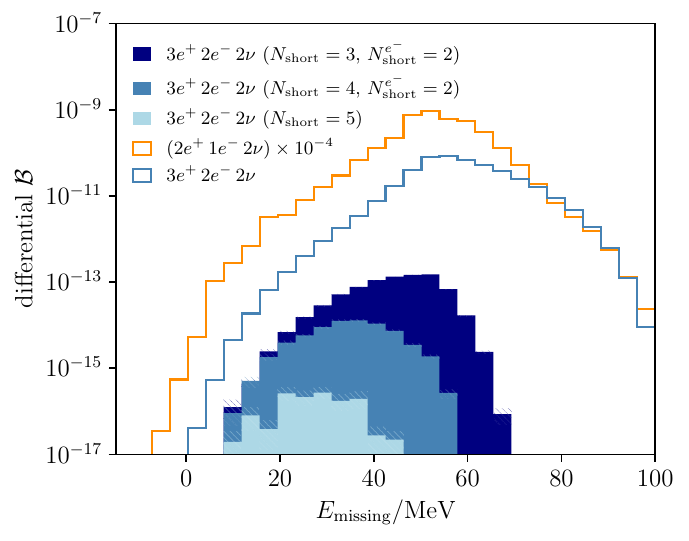}
    \caption{
    The reconstructed missing energy for $\mufiveeSMfull$ and $\muthreeeSMfull$ events in Mu3e. 
    For the former, the filled histograms show the subset of events with three, four, and five reconstructed short tracks, requiring exactly two reconstructed electrons.
    For the latter, we show the total rate downscaled by a factor of $10^{-4}$. 
    This points to a naive estimate of $3e$ events where one of the positrons undergoes Bhabha scattering to produce a four-track final state.
    \label{fig:missing_E_dists_eminus}  
    }
\end{figure}

The backgrounds to these three- and four-tracks selections will likely be larger.
For instance, mis-reconstructed $\muthreeeSMfull$ events will likely overwhelm the three-track channel due to the unavoidable missing energy.
We also note that three-track $5e$ decays are unlikely to be a background to the $\muthreeefull$ signal thanks to the two electrons and the significant amount of missing energy.
Nevertheless, the four-track $5e$ decays may very well be within the Mu3e reach.
The dominant background will likely be from $\muthreeeSMfull$ decays with a single Bhabha scatter.
Assuming the conservative numbers in \cref{sec:backgrounds}, this gives about $8\times 10^{6}$ events before any kinematical cuts.
This number is probably much smaller as the positrons from these decays are softer than Michel ones and will produce observable $e^+e^-$ pairs less often.
We show the missing energy distribution for different track requirements in \cref{fig:missing_E_dists_eminus}.
As a guide to the eye, we also show the $\muthreeeSMfull$ rate downscaled by $10^{-4}$, representing the rate for $3e$ events with an observable Bhabha pair.
The latter must be significantly suppressed to successfully measure the $5e$ decays through a four-track with two electrons signal selection.
Having identified the need for more flexible signal definitions to enhance the observable SM rate, we leave a detailed study of efficiencies and backgrounds to the collaboration.

\section{Conclusions}
\label{sec:concl}

The next generation of high-intensity experiments involving muons is poised to reach new levels of sensitivity to lepton flavor violation, down to $\mathcal{B}\sim 10^{-15}$. 
Typically, these experiments are designed with one particular physics goal and measurement in mind, such as $\mu^+ \to e^+e^+e^-$ decay in the case of the Mu3e experiment. 
However, the low-background environment, high intensity, and detector capabilities can turn these experiments into sensitive probes of dark sectors, putting them at the forefront of sensitivity to dark photons, dark scalars, or axion-like particles. 

In this work, we have analyzed the scenario of muon decay to 5 charged tracks, $\mufiveefull$. 
In the Standard Model, such a decay mode (accompanied by the emission of two neutrinos) is highly suppressed by powers of $\alpha$ resulting in a total branching ratio of $4\times 10^{-10}$.
The observable branching ratio will depend on the selection strategy, ranging from $\mathcal{O}(10^{-15})$ for five-short-track events to $\mathcal{O}(10^{-13})$ for events two observable electrons and two observable positrons.
In contrast, the higgsed $U(1)_1$ dark sector model we consider, generates a cascade of decays, resulting in a high multiplicity of electrons and positrons without further suppression by coupling constants.
Therefore, such dark sectors may induce the $3e^+2e^-$ decay signature with rates far above those of the SM backgrounds.
In particular, as shown in \cref{fig:missing_E_dists}, the SM decays do not pose a threat to $\mu \to 5e$ new physics searches since the branching ratio for the SM $\mu \to 5e2\nu$ decay with $E_{\rm missing} < 10$~MeV is of the order of $\mathcal{O}(10^{-15})$ or below.

The Mu3e collaboration currently does not have a dedicated study of a signal composed of five charged tracks. 
To that end, we have performed a simplified simulation of the Mu3e detector to determine a realistic signal efficiency. 
The requirement to have multiple energetic particles reduces the signal efficiency for the $\mu \to 5e$ signature down to the level of $\mathcal{O}(0.1\% - 1$\%), depending on the masses of the dark particles, indicating a resulting sensitivity to branching ratios of order $\mathcal{B} (\mufivee)\sim 10^{-12}$ in the most optimistic background-free case. 
For the dark sector masses in the tens of MeV range, the Mu3e sensitivity would translate to a probe of the lepton-flavor-violating muon-electron-dark Higgs coupling at the level $|y_{\mu e, e\mu}| < 10^{-13}$ and below, as shown in \cref{fig:sensitivity}. 
Recalling that such couplings originate from dimension-five operators, Eq. \eqref{eq:LFV_lagrangian}, we arrive at the future sensitivity of Mu3e to new physics scales $\Lambda$ as high as $\Lambda \propto 10^{15}-10^{16}$~GeV. 
We conclude by encouraging the Mu3e collaboration to perform a dedicated study of ``5e" physics channels, both in the neutrinoless new physics channel and in the SM channel, where neutrinos carry additional missing energy away.

\section*{Acknowledgements} 
We thank Drs.
B. Echenard, A.-K. Perrevoort, and F. Wauters for valuable discussions and correspondence.   
JZ and TM  acknowledge support in part by the DOE grant de-sc0011784 and NSF OAC-2103889. MP is supported in part by the DOE grant DE-SC0011842.
This research was supported in part by Perimeter Institute for Theoretical Physics. 
Research at Perimeter Institute is supported by the Government of Canada through the Department of Innovation, Science and Economic Development and by the Province of Ontario through the Ministry of Research, Innovation and Science.

\bibliographystyle{JHEP}
\bibliography{mu5e_biblio}
  
\end{document}